\documentclass{article}
\usepackage{graphicx} 
\usepackage{hyperref}
\usepackage[american]{babel}
\usepackage{csquotes}
\usepackage[style=apa,backend=biber,natbib=true]{biblatex}
\DeclareLanguageMapping{american}{american-apa}
\let\cite\autocite 

\makeatletter
\def\@fnsymbol#1{\ensuremath{\ifcase#1\or 1\or 2\or 3\or 4\or 5\or 6\or 7\or 8\or 9\else\@ctrerr\fi}}
\makeatother

\title{Two Means to an End Goal: Connecting Explainability and Contestability in the Regulation of Public Sector AI\thanks{This work is licensed under a \href{https://creativecommons.org/licenses/by/4.0/}{Creative Commons Attribution 4.0 International} license.}}
\author{Timothée Schmude\thanks{Corresponding author: \href{mailto:timothee.schmude@univie.ac.at}{timothee.schmude@univie.ac.at}}\, \thanks{University of Vienna, Faculty of Computer Science, Research Network Data Science, Doctoral School Computer Science}\, \textsuperscript{6}
	\and Mireia Yurrita\thanks{Utrecht University, Information and Computing Sciences Department}
	\and Kars Alfrink\thanks{Delft University of Technology, Department of Sustainable Design Engineering}
	\and Thomas Le Goff\thanks{i3, Télécom Paris, Institut Polytechnique de Paris}
	\and Sebastian Tschiatschek\thanks{University of Vienna, Faculty of Computer Science, Research Network Data Science} 
	\and Tiphaine Viard\textsuperscript{6}
	}
\date{}

\begin{document}

\maketitle

\begin{abstract}
	Explainability and its emerging counterpart contestability have become important normative and design principles for trustworthy AI as they enable users and subjects to understand and challenge AI decisions. 
	However, realizing these principles is difficult, as they assume different meanings in technical, legal, and organizational dimensions of AI regulation. 
	To resolve this conceptual polysemy, in this paper, we present the findings of an interview study with 14 experts to examine the intersection and implementation of explainability and contestability, and their understanding in different research communities. 
	We outline differentiations between descriptive and normative explainability, judicial and non-judicial channels of contestation, and individual and collective contestation action. 
	We further describe the main points of friction in the realization of both principles, including the alignment between top-down and bottom-up regulation, the assignment of responsibility, and the need for interdisciplinary collaboration. 
	Lastly, we formulate three recommendations for AI policy to implement both principles through a Regulation by Design perspective. 
	We believe our contributions can inform policy-making and regulation of these core principles and enable more effective and equitable design, development, and deployment of trustworthy public AI systems.
\end{abstract}


\section{Introduction}

Explainability and contestability are central principles in the development and deployment of trustworthy public AI systems~\cite{zuger_ai_2023}. However, while these principles are defined and discussed in both explainable AI (XAI)~\cite{lyons_algorithmic_2023, yurrita_disentangling_2023, hirsch_designing_2017} and legal~\cite{maxwell_meaningful_2023, kaminski2021right, almada_human_2019} research, an approach connecting these perspectives has not yet been adopted. The need for this unified approach is evident when considering the treatment of both principles in current AI regulation, such as the General Data Protection Regulation (GDPR) and Digital Services Act (DSA). While these texts can be understood to provide for explanations in AI systems to ensure contestability\footnote{The right to explanation is debated~\cite{wachter_counterfactual_2017, selbst_meaningful_2017, casey_2019}, but texts such as the GDPR, DSA, and EU AI Act provide for explainability in algorithmic decisions~\cite{maxwell_meaningful_2023}.}, they do not include guidelines to translate their legal provisions into tangible system requirements~\cite{maxwell_meaningful_2023}.

Explainability serves the purpose of increasing stakeholders' understanding of an AI system and to enable informed decision-making~\cite{langer_what_2021}, while contestability allows stakeholders to challenge and appeal algorithmic decisions~\cite{lyons_whats_2022}. A range of methods (`mechanisms'~\cite{Alfrink2022}) has been suggested in XAI and HCI through which these principles are realized, such as explaining feature importance~\cite{Ribeiro2016}, providing counterfactuals~\cite{wachter_counterfactual_2017}, or requesting human intervention~\cite{almada_human_2019}. But two main aspects remain underexplored: i) how explanation and contestation mechanisms intersect, and ii) how to proceed when implementing these mechanisms according to regulatory provisions. 

\subsection{Research questions}

We formulate two challenges that impede the implementation of 
explainability and contestability. First, both explainability and contestability are polysemic---they take multiple meanings depending on context---and require differentiation, as XAI, design, and legal research all employ the same terms but do not necessarily refer to the same concepts. Furthermore, the concrete realization of both principles depends on the involved actors~\cite{hoffman_explainable_2023}, domain~\cite{wang21}, and use-case setting~\cite{bove2022}. This multiplicity of meaning and realization excludes a one-size-fits-all approach~\cite{freiesleben2023, phillips_four_2021} and instead requires guidelines that can be applied in a variety of contexts~\cite{maxwell_meaningful_2023}. And second, AI in high-stakes scenarios is a comparably new phenomenon, with both theoretical background and regulatory oversight still in development. Consequently, there are few best practices and guidance that can aid in the implementation of contestability~\cite{alfrink2024envisioning, kaminski2021right, lyons_whats_2022} and interdisciplinary approaches to the creation of legislation have only begun to be mapped out~\cite{nahar_regulating_2024}. Closing this gap between regulation and implementation requires policy-making that is evidence-informed~\cite{mair_understanding_2019}, i.e., that is supported by research that bridges disciplines and provides empirical grounding. To this end, our work is guided by the following research questions:

\begin{enumerate}
    \item \textit{Conception}: How can we map the intersection between explainability and contestability of AI systems on a conceptual, design, and policy level?
    \item \textit{Implementation}: How can policy-makers support the implementation of explainability and contestability in AI systems? 
\end{enumerate}

Our findings highlight distinctions between descriptive and normative explainability, judicial and non-judicial contestation channels, and individual and collective contestation action. According to participants, its capacity for citizen empowerment is a key feature of connecting explainability and contestability; participants furthermore stressed this intersection as instrumental to both explainability and contestability's effectiveness and highlighted that both principles are not effective if the underlying policy governing a system's deployment follows values that are incommensurable with those of trustworthy AI.

This work contributes an empirically grounded differentiation of explainability and contestability, a description of the points of friction in their implementation, and a discussion embedding both principles into the larger research landscape. Our objective is to support an evidence-informed~\cite{carden_knowledge_2009, mair_understanding_2019} discussion to guide the implementation of policies surrounding the deployment of explainable and contestable public AI systems.  

\subsection{Methods used}

We conducted task-based interviews with 14 experts in the regulation and design of AI systems. Participants were recruited from European universities and public institutions.\footnote{To allow participants the freedom of anonymous expression in the interviews, demographic as well as occupational details on their persons were omitted.} These participants were selected for their professional background (e.g., involvement in the EU institutions, standardization bodies, or legislative bodies), their knowledge of legislation and implementation of AI policies and their technical, legal, or design expertise. As this study aims to elicit details of how explainability and contestability are discussed and understood in policy-making circles--- a form of policy work that is not documented in academic publications~\cite{yang_future_2024}---we focused on sourcing insight from policy experts directly, rather than conducting 
a literature review. For participant recruitment, we relied on 
the authors' networks, snowball sampling, and direct invitations. The sample size followed qualitative research guidelines, focusing on code and meaning saturation~\cite{hennink_code_2017}.

The interviews consisted of three key elements: a card sorting activity, a use case discussion, and the exploration of a citation network. Participants first sorted 40 cards containing explanation and contestation mechanisms into self-defined categories, indicating their understanding of both principles. Participants then read a description of the French CAF\footnote{Caisse d'allocations familiale} welfare fraud detection algorithm~\cite{lighthouse_CAF, lemonde_CAF} and reflected on the use of explanation and contestation from the perspectives of a fictional welfare beneficiary who was flagged for fraud and the social security agency. This use case was chosen as it resembles a broader set of algorithmic decision-making systems used in public institutions~\cite{european_parliament_understanding_ADM_2019}. Lastly, participants explored an interactive network representation of academic literature on explainability and contestability.~\footnote{The graph is available under \href{http://contest.graphuzu.fr/}{contest.graphuzu.fr}.}

Interviews were recorded, transcribed, and analyzed using inductive thematic analysis~\cite{braun_using_2006} to examine participants' conceptual understanding, experiences with the implementation of AI regulations, and their perspectives on the involved stakeholders and collaboration between disciplines. 

\subsection{Structure of the paper}

The remainder of the paper is structured as follows: In Section~\ref{sec:background}, we give context for why the two principles of explainability and contestability matter for AI policy and introduce the concept of Regulation by Design~\cite{prifti_regulation_2024}. In Section~\ref{sec:intersection} and~\ref{sec:practice} we report on our study findings with respect to the intersection of both principles and their implementation in practice. In Section~\ref{sec:discussion}, we formulate these insights into recommendations to realize explainability and contestability through the lens of Regulation by Design. We give the study's limitations in Section~\ref{sec:limitations} and conclude in Section~\ref{sec:conclusion}.

\section{Why connecting explainability and contestability matters for AI policy}
\label{sec:background}

Explainability and contestability are two means to an end goal~\cite{maxwell_meaningful_2023}. This end goal is the development and deployment of trustworthy AI systems that preserve decision subjects' fundamental rights~\cite{floridi_establishing_2019}, support human agency and oversight~\cite{crootof_humans_2023}, and adhere to the principles of procedural justice such as transparency and outcome control~\cite{lee2019procedural}. However, implementing the high-level principles 
can prove challenging in practice~\cite{madaio_co-designing_2020, ferrario_how_2022}, which is why fostering a perspective that contains both policy and design considerations is essential. Because of the interdependence of legal frameworks, design practices, and institutional governance, connecting explainability and contestability in both policy and design is an essential step in realizing Regulation by Design. In the following, we situate our work in relation to public sector AI (Section \ref{sec:background_public_ai_systems}) and describe the concept of policy knots in AI systems design and use (Section \ref{sec:unknotting}). Finally, we introduce a definition of regulation and the dimensions in which it is realized, situating our work within the Regulation by Design framework (Section \ref{sec:regulation-by-design}). 

\subsection{Trustworthy AI systems in the public sector}
\label{sec:background_public_ai_systems}

AI systems deployed in public institutions can significantly impact individuals' fundamental rights, safety, or health~\cite{hupont_documenting_2023}. Research shows that these systems are frequently dysfunctional~\cite{raji_fallacy_2022, kearney_2024}, discriminatory~\cite{chouldechova2017}, and harmful through aggravating power imbalance~\cite{eubanks_2018, Nedzhvetskaya_2024} and restricting autonomy~\cite{prunkl_human_2022}. For these reasons, both researchers~\cite{baeza-yate_2022, Rudin2019, bogiatzis-gibbons_2024, huang_2024} and policy makers~\cite{high_level_expert_group_trustworthy_ai, phillips_four_2021, european_commission_laying_2021, oecd_recommendation_2019} have advocated that high-risk AI systems should align with value frameworks such as \textit{trustworthy} or \textit{responsible} AI~\cite{thiebes_trustworthy_2021, floridi_ai4peopleethical_2018, coeckelbergh2020ai}, which emphasize human agency, oversight, transparency, accountability, and fairness~\cite{floridi_establishing_2019}. Explainability and contestability support these frameworks by enabling people to understand~\cite{langer_what_2021} and challenge~\cite{henin_beyond_2022} AI decisions. Although both principles are integrated into various design frameworks~\cite{alfrink2024envisioning, langer_what_2021, hoffman_explainable_2023, lyons_algorithmic_2023}, their implementation as part of the EU AI regulation remains challenging~\cite{maxwell_meaningful_2023, nahar_regulating_2024, figueras_trustworthy_2021}.

\subsection{Unknotting explainability and contestability in policy,  regulation, and design}\label{sec:unknotting}

Our work in this paper follows an `unknotting' process, as we seek to understand explainability and contestability from a multifaceted viewpoint that combines policy, design, and technology perspectives. The `policy knot' is a metaphor that describes the complex interdependencies of technology, practice, and policy in the context of computing systems research.~\citet{jackson_policy_2014} describe it as ``the multiple gatherings and entanglements through which 
worlds of design, practice and policy are brought into messy but binding alignment''~\cite{jackson_policy_2014}. The policy knot concept has been used, for example, to study how the values embedded in AI-enabled creativity tools interact with the wider creative industry, and to describe acts of critique and resistance by users as a form of `unknotting'~\cite{shelby_creative_2024}. Similar analyses have been conducted for topics such as
    copyright implications of GenAI~\cite{centivany_mining_2024},
    automated grading systems~\cite{figueras_doing_2024},
    rejection of AI tools in user experience design practice~\cite{cha_understanding_2025},
    and
    online community platform policies regarding AI-generated content~\cite{lloyd_ai_2025}.

\subsubsection{From trustworthiness, through accountability and transparency, to explainability and contestability}

The EU Commission's HLEG's Guidelines for Trustworthy Artificial Intelligence define the property of trustworthiness as being (1)~lawful, 
(2)~ethical, 
and (3)~robust (both from a technical and social perspective)~\cite{high_level_expert_group_trustworthy_ai}. AI systems in this case are understood to be sociotechnical
~\cite{wenzelburger_algorithms_2022} and the trustworthiness in question should be ensured throughout systems lifecycles
~\cite{dhanorkar_who_2021}. From the seven key requirements of trustworthy AI, we focus in the following on transparency and accountability to describe their connections to explainability and contestability.

The HLEG describes \textit{transparency} as providing information about data and models~\cite{high_level_expert_group_trustworthy_ai}. This requirement is further decomposed into traceability (maintaining records of data and processes that lead to discrete outputs), communication (ensuring users know whether they interact with a machine or a human), and, most notably for our purposes here, \textit{explainability}, which means providing an account of the technical and human processes that went into an AI system's decision. In the case of significant impact, \textit{decisions} must be \textit{explainable} ``to the extent possible'' to those directly and indirectly affected (i.e., `decision subjects'), and must cover not only information about an individual decision (\textit{local} explainability) but also about the organizational processes, design choices, and rationale for deploying it (\textit{global} explainability.) If a model's output cannot be explained due to technical constraints (as is the case with black-box models~\cite{Rudin2019}), then other measures are required (e.g., traceability, auditability). 

The HLEG further describes \textit{accountability} as a way to adhere to the principle of fairness by ensuring actors take or are assigned responsibility for systems and outcomes throughout their lifecycle. This requirement is operationalized through auditability, the minimization and reporting of negative impacts, and \textit{redress}. Redress means providing `accessible mechanisms'~\cite{pi_toward_2025} to remedy an undesirable or unfair impact, which we describe as individual, local \textit{contestability}. 

Using this terminology we can draw out a basic connection between trustworthiness, transparency, explainability and contestability: Trustworthiness depends among other things on systems being \textit{transparent}. This transparency is realized by making sure AI model outputs and the decisions are \textit{explainable} to decision subjects, which provides them with the information required to \textit{contest} a decision by using \textit{redress} mechanisms. Based on these considerations, we provide two working definitions of explainability and contestability.

\subsubsection{Explainability}

Explainability means providing information about an AI system's logic, core parameters, and specific purposes. `Good' explanations are expected to adapt to stakeholders' expertise and objectives~\cite{miller_explanation_2019, langer_what_2021}, be it evaluating fairness, understanding the deployment context, or contesting a decision~\cite{deck_critical_2024, schmude2024information, alfrink2024envisioning}. 
Although explainability is provided for in many policy texts~\cite{DSA_EU_2021, GDPR2016a, Platform_Workers_Directive}, these texts leave implementation details unspecified, such as whether to use global or local explanations and whether to provide them before or after decisions~\cite{maxwell_meaningful_2023}. 

\subsubsection{Contestability}

Contestability describes how various stakeholders, from human operators to decision subjects and third parties, can 
challenge algorithmic decisions~\cite{Alfrink2022, lyons2022_conceptualising, almada_human_2019}
through data input control, decision revision requests, and various audit methodologies~\cite{shen_everyday_audits_2021, lee2019procedural, lyons_algorithmic_2023}. In both design and policy frameworks, the principle of contestability is described to be enabled through explainability~\cite{almada_human_2019, maxwell_meaningful_2023}. Contestation rights are contained in the EU Charter of Fundamental Rights~\cite{borraccetti_2011}, in GDPR Article 22, and the Council of Europe's Convention 108+\footnote{The Council of Europe's Convention for the Protection of Individuals with regard to Automatic Processing of Personal Data.}~\cite{almada_human_2019, CoE_Convention_108+}.

\subsubsection{Limitations of explainability and contestability}

Current limitations to explainability and contestability implementation in AI systems stem from both 
technical constraints and 
policy challenges.

From a \textit{technical} perspective, creating workable forms of explainability is not just technically challenging, but even impossible for AI systems~\cite{keller_exclusivity_2021} 
whose internal workings are fundamentally opaque to outside human observers~\cite{raees_explainable_2024}, i.e., black box models~\cite{Rudin2019}. 
Additionally, without defined benchmarks for explainable AI, user evaluation methods are prone to many knowledge gaps depending upon the varying explainability needs of different users~\cite{raees_explainable_2024}.
Contestability faces the fundamental challenge that much of the current AI landscape is tailored to `static AIs', i.e., systems that do not allow humans to interactively interrogate AI decisions or to update their decisions in response to user input~\cite{leofante_contestable_2024}. 

From a \textit{policy} perspective, explainability demands a workable reconciliation between explainability's technical and commercial limitations and an array of public governance standards~\cite{keller_exclusivity_2021}. Commercial limitations include 
the reluctance of tech firms to disclose trade secrets or other confidential information through explainability as well as the increasing costs 
of adhering to AI regulation 
\cite{keller_exclusivity_2021}.
Contestability suffers from governance processes that are 
closed to effective public contestation~\cite{keller_exclusivity_2021} and furthermore requires that policies 
openly specify the regulations and determinations of adequacy for specific applications. These include proper attribution of accountability and effective methods of compensation when decisions are challenged~\cite{aler_tubella_contestable_2020}.

\subsection{Regulation by Design: from regulatory intent to institutional governance}
\label{sec:regulation-by-design}

This section serves to define the process of regulation and to introduce the key concept of Regulation by Design (RBD), where regulation is a multi-layered approach encompassing both  regulators and regulated actors~\cite{prifti_regulation_2024}. Here, regulation is a process involving the ``sustained and focused attempt to alter the behavior of others according to defined standards or purposes'' to produce a broadly defined outcome~\cite{black2001}. 

While classical regulation relies on norms imposed by public institutions (`standard-setting' and `command-and-control' mechanisms~\cite{breyer_regulation_2009}), modern regulation evolves towards the concept of `Regulation by Design'~\cite{Morgan_Yeung_2007, yeung_2015}. RBD aims to embed regulatory objectives directly into technological architectures, making design choices a vehicle for achieving social governance of systems~\cite{prifti_regulation_2024}. In this conception, regulation is understood as a rule-making activity performed through social practices. 
Design is understood as more than a tool: a social practice situated in-context that makes changes in the environment. 
Unlike the law or social norms which rely on incentives and sanctions, design can disable the possibility of non-compliance, making it a uniquely powerful form of regulation. This conceptualization addresses legitimacy concerns by allowing for greater accountability for design practices that might bypass democratic processes. 

\begin{figure}[htb]
    \centering
    \includegraphics[width=1.0\linewidth]{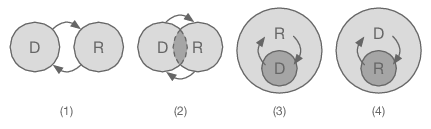}
    \caption{We see four possible ways that the relationship between the design (D) and regulation (R) of technology can be conceptualized: (1) design and regulation are independent spheres, (2) design and regulation partially overlap, (3) design is embedded inside of regulation, or (4) regulation is embedded inside design. In all cases, we assume an interactional relation between the two spheres. Technology itself, the product of design, and the object of regulation, is not depicted.}
    \label{fig:relations-between-design-and-regulation}
\end{figure}

Regulation by Design corresponds to type 3 in Figure~\ref{fig:relations-between-design-and-regulation}: design is embedded inside of regulation and is one of a variety of ways in which regulation makes changes in the world.
This view of design practice has concrete implications for regulation: regulation must accommodate evolving technological capabilities, societal values, and ethical principles while avoiding obsolescence. This is made more difficult by the fact that modern regulation relies on regulated actors to operationalize abstract concepts through co-regulation, a collaborative mechanism where stakeholders---public authorities, developers, and end-users---translate legal norms into actionable practices. This dynamic creates a feedback loop where regulatory intent (`the spirit of the law') and operational realities continuously influence each other~\cite{prifti_theory_2024}.

Legal frameworks, design practices, and institutional governance are thus interdependent, it is essential to connect explainability and contestability in policy and design to realize RBD. This means finding congruent terminologies that work between disciplines, and fostering a deeper understanding of the problems that policy-makers and designers face in their work with these principles---a main concern of this paper.

\section{Understanding the intersection of explainability and contestability of AI systems}
\label{sec:intersection}

This section addresses our first guiding question (RQ1) by mapping the intersection of explainability and contestability of AI systems with respect to their conceptual, design, and policy dimensions. We use evidence from the task-based interviews and quotes from the participants (referred to as ``P'') to differentiate both principles and list their goals, mechanisms, and challenges. We first map how participants defined explainability and contestability (Section \ref{sec:results_polysemy}) to draw out differences in their understanding. We then show that a consensus emerged among participants on the fact that the two concepts follow the same goals while using different mechanisms (Section \ref{sec:results_intersection_xai_conai}). However, the intersection between the two notions is complicated by disciplinary gaps and divergences (Sections \ref{sec:results_gaps} and~\ref{sec:practice}).

\subsection{Differentiating explainability and contestability}
\label{sec:results_polysemy} 

\textbf{Explainability} is connected to two main notions: 1) understanding the technical workings of an AI system, such that ``\textit{the human user [...] is not treating anymore the machine as an oracle}'' (P2); and 2) understanding the norms and reasons governing the AI's decisions, deployment, and institutional embedding, such as ``\textit{know[ing] who are the people in charge or who I can contact to give more information}'' (P1). We define the first kind as \textit{descriptive explainability} and the second kind as \textit{normative explainability} or \textit{justifiability}. 

\textbf{Contestability} means allowing stakeholders to challenge AI decisions, such as enabling regulators to ``\textit{critically process the information provided to them and push back against it}'' (P6) and affected persons to ``\textit{understand the situation and to file complaints}'' (P6). We define two key distinctions in how contestation is realized: The first distinction is between \textit{collective action} and \textit{individual action}. While individual contestation allows decision subjects to challenge AI decisions that affect them, collective contestation means joining with other stakeholders to mount a more general and broader contestation effort, i.e.,``\textit{translating personal issues into general matters and public fights}'' (P12). The second distinction is between \textit{judicial channels} and \textit{non-judicial channels} of contestation. Judicial channels use formal means provided by the judicial system to contest decisions in court, colloquially described as ``lawyering up'' (P9), while non-judicial channels support issue resolution through design solutions or direct human intervention, such as mediation systems or ombudspersons. Differentiating both the the type of contestation action and the channels used is essential to pinpoint the meaning of contestability: ``\textit{For legal scholars, [...] there is this idea of centering judicial proceedings and centering the courts, even though most of what we could call the `life of the law' is not usually in the courts}'' (P9).

\subsection{Explainability and contestability follow the same goals but use different mechanisms}
\label{sec:results_intersection_xai_conai}

\subsubsection{Goals} Explainability and contestability work towards the same goals and purposes. These goals can be summarized as supporting the rule of law and empowering citizens by alleviating opacity and information asymmetry: ``\textit{We cannot contest [if] we don't know what is used and to what purpose and how it works}'' (P3). 

Explanations serve these goals through the provision of descriptive and normative information, which in turn enables the assessment of two distinct aspects: the acceptability of individual decisions and the justifiability of a system's deployment overall. This dual assessment, in turn, prompts actions of contestation at the individual or collective level.  The acceptability of individual decisions becomes crucial when subjects face unfavorable outcomes and must decide whether to contest them. P9 noted that acceptability is important so that citizens would not use contestation channels only to impede the process, i.e., to ``\textit{not just to throw a cog in the wheel and delay procedures that you know that are going to be inconvenient to you, but are otherwise acceptable}'' (P9). When individual contestation is not possible, explanations can enable collective contestation of the entire AI system if, e.g., the underlying policy is identified as unfair or dysfunctional. Explanations thus enable stakeholders to assess their attitude in relation to an AI system both locally and globally and to find corresponding contestation actions and channels. This differentiation is crucial, as it determines whether contestation is mounted between the decision subject and an officer of the institution, or between a collective of decision subjects and the institution as an organization.

\subsubsection{Mechanisms} Realizing the principles of explainability and contestability in actual application contexts requires processes or techniques that implement them. We describe these processes and techniques as \textit{mechanisms}.

\begin{figure}[htb]
    \centering
    \includegraphics[width=1.0\linewidth]{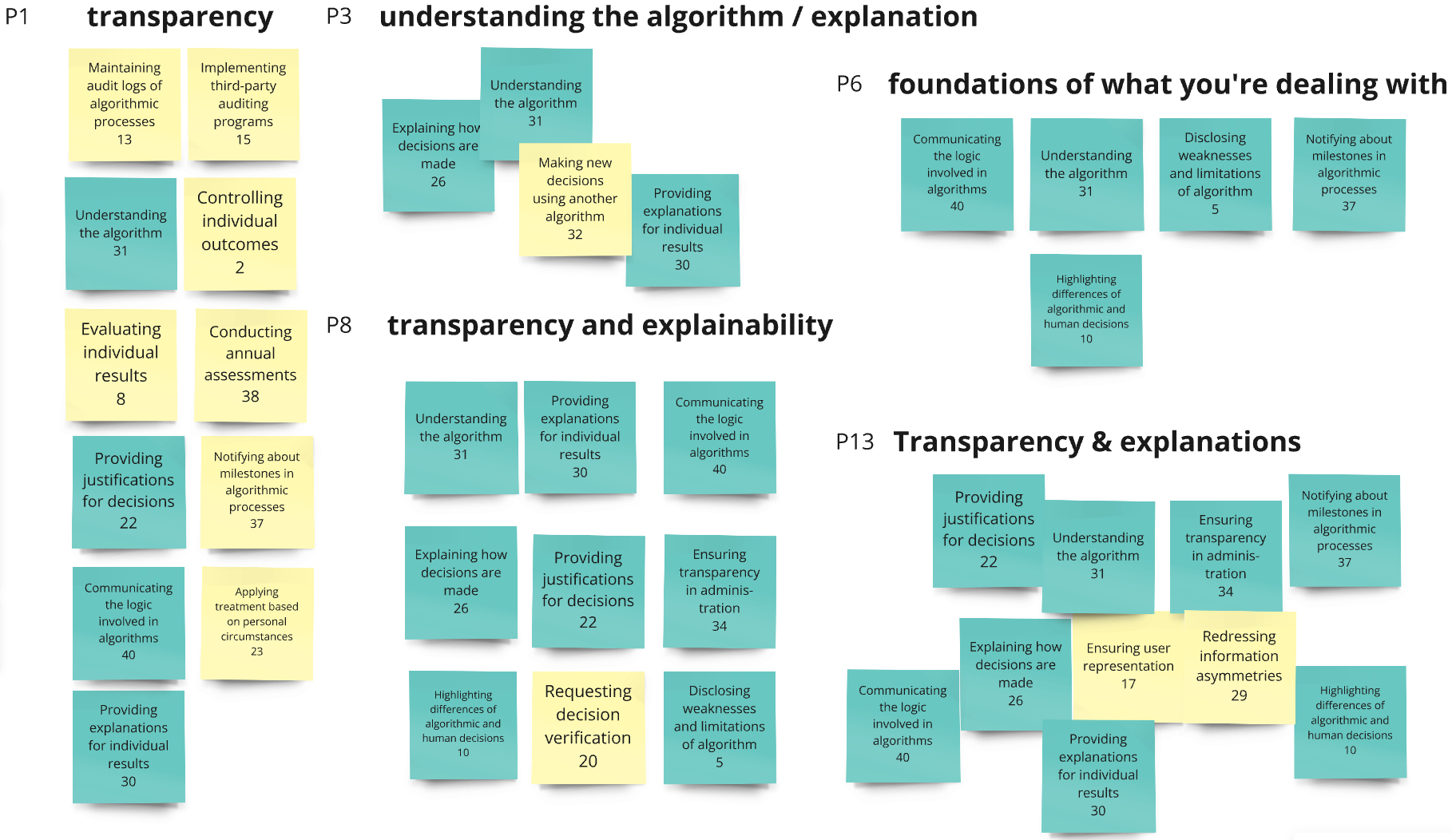}
    \caption[Card sort explainability clusters]{\textbf{Card sort explainability clusters.} Six different card sorts that all contain cards that participants related to transparency or explainability. Often, participants created exactly one cluster with a similar name to those shown. Cards that can be found in multiple of the six clusters are colored petrol, cards that are only in one cluster are colored yellow.}   
    \label{fig:card_sort_explainability}
\end{figure}

\begin{figure}[htb]
    \centering
    \includegraphics[width=1.0\linewidth]{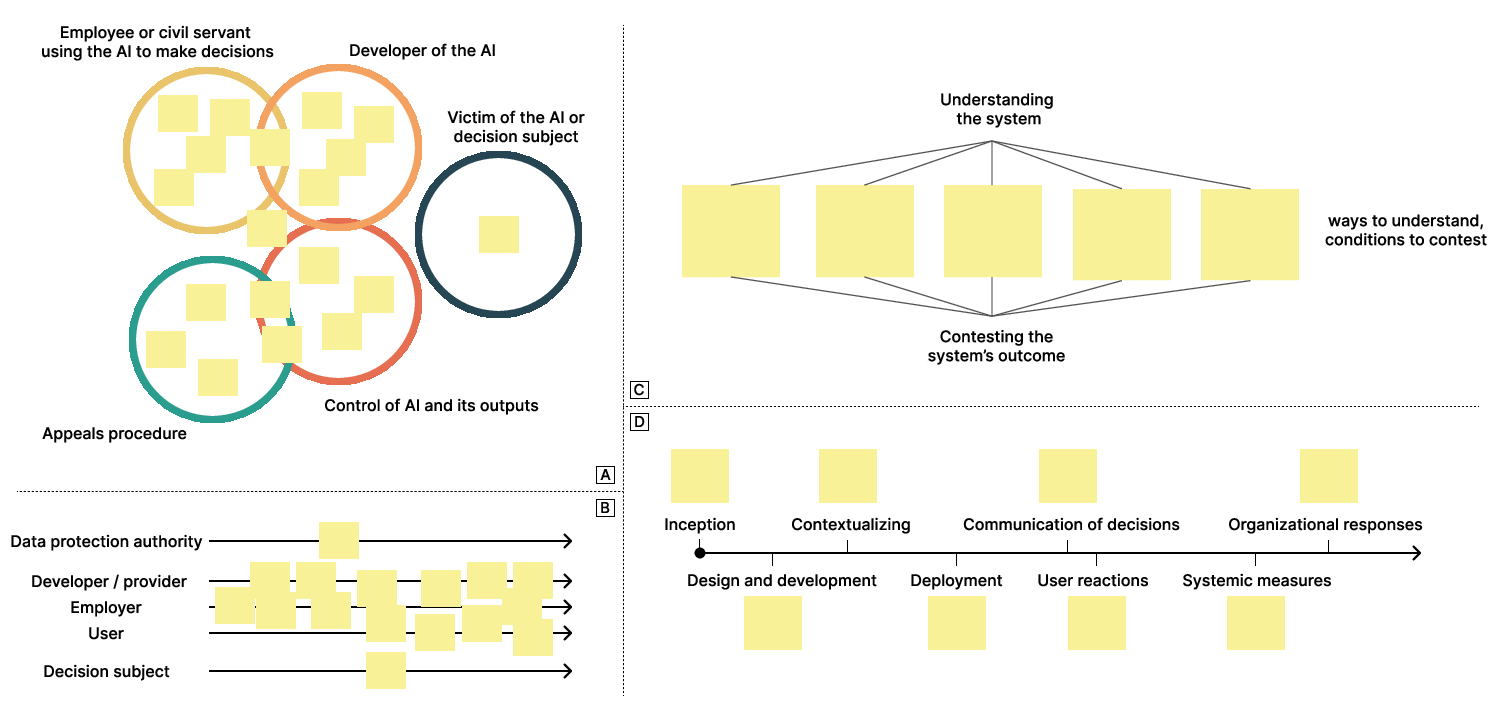}
    \caption[Card sort structures]{\textbf{Card sort structures.} Depicted are the structures of four completed card sorts. Participants chose different criteria and dimensions to sort the cards into clusters, including spheres of responsibility (A), responsibility over time and per role (B), ways in which mechanisms connect both to understanding and contesting (C), and an allocation to the implementation process over time (D).}
    \label{fig:card_sort_structures}
\end{figure}

In the interviews, participants were presented with a selection of cards listing explainability and contestability mechanisms and were asked to sort them into clusters. For explainability, participants created clusters that were titled with `transparency', `understanding the system', `explanations', or `foundations of what you're dealing with'. A selection of examples is depicted in Figure~\ref{fig:card_sort_explainability}. In contrast, contestability was perceived as a less uniform principle, comprising clusters titled `control', `appeals procedure', `rectification', `judicial remedies', and `auditing'. The variety of these clusters shows that participants understood the principles differently depending on the involved actors, subjects, goals, channels, and the multiplicity of ways to realize them. This variety is illustrated in Figure~\ref{fig:card_sort_structures}.

\begin{figure}[htb]
    \centering
    \includegraphics[width=1.0\linewidth]{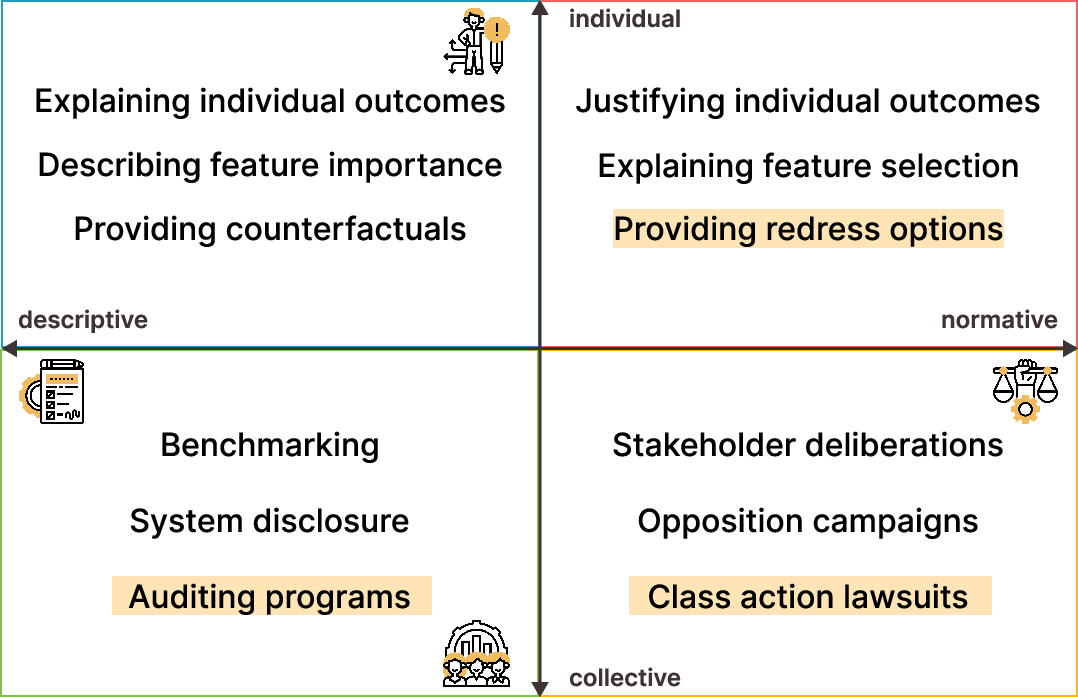}
    \caption{\textbf{Mapping mechanisms of explainability and contestability.} Depicted is a grid that embeds three dimensions of explainability and contestability: descriptive versus normative approaches (x-axis), individual versus collective action (y-axis), and judicial (highlighted in yellow) versus non-judicial channels. Each quadrant contains examples of mechanisms that are assigned to the corresponding dimensions. To illustrate, we describe two mechanisms: `Explaining individual outcomes' means providing decision subjects (individual) with a description of the algorithmic output leading to their decision (descriptive). `Stakeholder deliberations' can be used to align the values of an AI system with those of the citizens (normative), thus providing a form of democratic control (collective).} 
    \label{fig:duality_ex_and_con}
\end{figure}

Employing the differentiation introduced in Section~\ref{sec:results_polysemy}, we propose to map mechanisms of explainability and contestability along three main axes: Whether they provide descriptive or normative information, whether they support individual or collective contestation, and whether the mechanism employs judicial or non-judicial channels. A visual representation of this mapping is shown in~\autoref{fig:duality_ex_and_con}.

\subsubsection{Challenges} Explainability and contestability have their limits in alleviating opacity and information asymmetry. Participants articulated three circumstances in which explanation and contestation mechanisms would be ineffective: \\ i) the system's purpose is to enforce sanctions on decision subjects (i.e., the system is ``\textit{badly designed on purpose}'' (P4)), ii) the sampling strategy targets vulnerable populations disproportionally, and iii) non-judicial channels of contestation are obscured or not available, i.e.,``\textit{you cannot even exercise [the right to contestation] properly without assistance}'' (P9). In consequence, individual explanation and contestation action cannot empower citizens if the policy embedded in an AI system follows values that do not adhere to those of trustworthy AI. 

Participants further problematized that individual contestation primarily serves individual interests and has less potential to effect change on a systemic level. Decision subjects ``\textit{want to fight for \texttt{their} privacy and \texttt{their} freedom. It's very different from fighting for privacy with a big P}'' (P12). Collective contestation was posited as a more effective alternative for ``\textit{contesting decision patterns [...] through things like class actions}'' (P9), but enacting this form of contestation was perceived to be insufficiently explored within the current EU jurisdiction. 

Inciting political debates on a collective level was further seen as a capacity of contestability and normative explanations (justifications) rather than descriptive explanations. A descriptive explanation does not justify that ``\textit{the decision is correct, accurate, and legit}'' (P11). Descriptive explanations thus lack the normative foundation that is essential for contestability, which highlights the need for justifications as way of providing information on a system's legitimacy and value alignment that can effectively support contestation. 

Lastly, preventing a surge of contestation requests and the gaming of decision processes were perceived as prevalent challenges when implementing explanation and contestation measures: ``\textit{If [the institutions] do it properly, they will be submerged by requests and contestation}'' (P12). Both excessive contestation and attempts to game the system stem from citizens' intention to take control over the decision process to achieve favorable results, especially if they feel that their values are not adequately represented. Participants stressed that when confronted with a badly justified algorithmic decision, citizens ``\textit{adapt their behavior to their [antagonistic] interpretation of the algorithm}'' (P5). As a remedy, decision subjects could be included in the design process of public AI systems to incorporate their perspectives, improve transparency in rationale, increase acceptance of unfavorable decisions, and fulfill administrative requirements ``\textit{to consult the population in terms of impact and gather public feedback}'' (P9).

\subsection{Gaps between disciplines impede mapping the intersection of explainability and contestability} 
\label{sec:results_gaps}
While the ``right to explanation''~\cite{casey_2019, selbst_meaningful_2017, wachter_counterfactual_2017} and judicial ways of contestation, such as appeals and redress, were well known to the interviewed experts, the term ``contestability'' and the corresponding body of work in design and HCI research were new to some.
Participants commented on this, stating ``\textit{I'm thinking in which world I was living [...], I didn't notice that [contestability] was so well-developed}'' (P3) and ``\textit{I'm not familiar with the concept of contestability as such, [...] I rather use `redress', for example}'' (P10). While this points to differences in vocabulary, it also indicates a more conceptual lack of connection between fields. Participants confirmed this gap when exploring the citation graph and stated that the lack of connection between research fields aligned with their experience: ``\textit{My experience is, in fact, that the different disciplines are not talking to each other.}'' (P8) and ``\textit{You have some [communities] that are closer, like [...] legal people sometimes go to the technical part, [...] but some others are not really talking}'' (P9). An absence of connection to the legal literature was especially noticed in relation to design, sociotechnical, and ethics literature. Thus, this definitional and conceptual separation between the disciplines, as well as the conceptualization of the relationship between design and regulation, are main challenges in mapping the intersection between explainability and contestability.

\section{Explainability and contestability in practice: points of friction in regulatory, institutional and technical implementations}
\label{sec:practice}

This section addresses our second research question (RQ2) by addressing points of friction in the implementation of explainability and contestability. We use evidence from the interviews to show that their implementations need to cohere to the spirit of the initial concepts (Section \ref{sec:spirit_of_law}), that their translation from theory to practice is a shared responsibility among different ``regulators'' (Section \ref{sec:responsibility}), and that implementations profit from collaborations between communities (Section \ref{sec:painful}).

\subsection{Implementations of explainability and contestability should keep in mind the spirit of the law}
\label{sec:spirit_of_law}

Participants identified a key concern of implementing regulatory provisions in the question of whether these implementations would capture the regulation's intent, i.e., the \textit{spirit of the law}. Participants who were members of the EU AI HLEG described that the group asserted in unison the importance of explainability in the AI Act: ``\textit{lawyers agreed in there, and human rights experts agreed in there, practitioners, [...] there was no doubt that this is a fundamental requirement}'' (P8). Explainability was then integrated into the principles and the seven key requirements even though its prospective implementation was already registered as an issue: ``\textit{[W]e put as principle something that wasn't really possible 100\%. But we felt that it was important to have this because it also reflects on [...] contestability}'' (P10). As such, requirements such as those stated in the AI Act are subject to several transformations before being realized in practical applications: they are formalized in legal texts and technical standards, integrated into national jurisdiction, and only then implemented in public institutions. In the course of this process, preserving the spirit of the law is not a given.

In particular, participants feared that downstream applications would not keep the intended safeguards intact due to diverging interpretations. P8 described their experience when meeting lawyers who were ``\textit{discussing whether a software application could be a high-risk application. [...] I thought, no, this was not the intention, [it was] to safeguard certain principles}'' (P8). Another participant stated that ``\textit{it's not just about complying to the letter of the EU AI Act but also the spirit, the spirit of the act is to empower affected individuals to safeguard their rights}'' (P11). When understanding design as a vehicle of regulation, then embedding the legislation's intention and corresponding values into design is crucial, rather than designing for compliance ``to the letter''. We propose one method of value alignment in Section~\ref{sec:recom_1}.

\subsection{Implementing explainability and contestability clarifies how responsibility moves between regulators}
\label{sec:responsibility}

In the interviews, participants repeatedly considered which actors would be responsible for implementing explanation and contestation mechanisms (these considerations are visually depicted in Figure~\ref{fig:card_sort_structures} A and B). Participants considered `regulators'~\cite{prifti_regulation_2024}, including policymakers, standardization bodies, data protection authorities, and developers, to be the main actors that shared the ``\textit{responsibility to ensure user representation in the development and the use of the AI}'' (P5). Technical standards were perceived to be one of the main components to clarify the allocation of responsibility, but their enforcement raised questions. Due to the conceptual polysemy with which legal texts treat both design principles, fulfilling their respective responsibility means that regulators are forced to interpret the provisions, potentially resulting in conflicting points of view between the `executive' and `organizational' levels. 

Importantly, participants also stated that the regulation of AI systems has not yet actually taken place, as ``\textit{nobody has implemented the EU AI Act yet [...]. There's no national law to set down sanctions}'' (P6). This has two implications: While first, the task of interpreting the regulatory provisions is not clearly assigned between EU jurisdiction, national authorities, and public institutions, second, the process of assigning responsibility for this interpretation can still be shaped, leaving room to delineate ``\textit{how to handle conflicts}'' (P6) and ``\textit{how we are going to adapt our legal system}'' (P1). We outline ways to create a composite form of responsibility through indirect and direct control mechanisms in Section~\ref{sec:recom_2_and_3}.

\subsection{Collaboration between communities strengthens the implementation of explainability and contestability} \label{sec:painful}

Participants who had come into contact with both legal and design research on AI regulation regularly highlighted the potential (and current shortcomings) of interdisciplinary collaboration. Participants criticized that technical explanations of AI behavior often were not available ``\textit{only because at the beginning of the process, they haven't thought about that}'' (P1). In consequence, and because ``\textit{explainable AI does not fit into the justification of legal decisions}'' (P2), policy-makers were considered to have an incomplete picture of technology. Explainability was described to facilitate communication between disciplines, since ``\textit{as soon as we start to explain what we are doing, [...] everyone else, also from different disciplines, can understand}'' (P8). In turn, participants proposed that legal studies could improve the normative force of design research by giving it ``\textit{a bit more punch}'' (P9):

\begin{quote}
    \textit{[O]ne thing that legal studies can help is to say: `No, you have to care about this not just because we are a bunch of hippies trying to save the world, but also because if you don't, you're going to have lots of problem with the law [...]  or even have your system not being commercialized in a particular jurisdiction.} (P9)  
\end{quote}

While the advantages of interdisciplinary collaboration thus become evident, following through on it was described to be ``\textit{painful}'' (P8). Especially for people removed from the technical sides of AI, ``\textit{even explaining the concept of explainability sometimes can be challenging because they have to understand that AI is a black box}'' (P14). Similar comments were made about non-judicial contestation actions, which were not considered to be well-known in legal disciplines (we formulate recommendations for the implementation of non-judicial channels in Section~\ref{sec:recom_2_and_3}). Participants who had experience in both technical and legal disciplines described it as ``\textit{very challenging}'' to ``\textit{find a right level of abstraction where we don't get too bogged down on the details [...] versus where we don't generalize too much}'' (P9). 
Finding methods that facilitate shared conceptual understanding and vocabulary is thus essential to effectively inform policy and implementation.

\section{Perspectives for AI policy}
\label{sec:discussion}

Drawing from the insights generated in the interviews, we suggest recommendations towards the implementation of explainability and contestability through the lens of Regulation by Design~\cite{prifti_theory_2024} (Section \ref{sec:regulation-by-design}). To this end, we organize our recommendations in two main groups: (1)
those in which design practice contributes to regulation and (2)
those in which internal governance systems are used as mechanisms that steer practices in public institutions.  

\subsection{Regulating, towards explainability and contestability, by design and by law} 
\label{sec:recom_1}

We established in Section~\ref{sec:regulation-by-design} that Regulation by Design views regulation as a rule-making activity performed through social practices like design. As design can simply disable non-compliance, it is a powerful form of regulation. We, therefore, advocate for greater accountability for design practices that might bypass democratic processes.

\subsubsection{Recommendation 1: Making design decisions open to public deliberation.}
The design of AI systems involves encoding legislation into software~\cite{Zouridis2020}, meaning that design decisions about input features, data types, and human-AI interactions become policy decisions that are no longer delegated to public deliberation but rather to third-party developers~\cite{mulligan2019procurement}. Because important decisions are made early in the design process, participants emphasized the need to involve stakeholders, particularly decision subjects, through ``early-stage deliberations''~\cite{kawakami2024}. By integrating contestability as a co-designed and technical feature of the system itself rather than as a legal standard to meet, public institutions could align values embedded in AI systems with those of citizens. This might improve the acceptability of AI decisions while avoiding excessive contestation during operation. To this end, we highlight two main aspects of participatory approaches: i)~the deliberations' level of abstraction and ii)~the participation mechanisms. Regarding the level of abstraction, prior work suggests that, instead of focusing on technical design decisions, participatory approaches should center around the values and policies embedded in code~\cite{abdu_2024}. These values and policies can, in turn, be selected through citizen-wide participation before being embedded in AI systems~\cite{bogiatzis-gibbons_2024}. Regarding participation mechanisms, citizen assemblies or advocacy groups that represent groups that are unable to participate fully (e.g., children) can be alternatives to direct participation mechanisms~\cite{bogiatzis-gibbons_2024}.

\subsection{Strengthening the intersection of explainability and contestability through internal AI governance mechanisms}
\label{sec:recom_2_and_3}

Internal governance systems coordinate social action through formal and informal mechanisms. Governance acts as a meta-regulative activity steering how other practices, like design, regulate. We suggest that internal AI governance systems strengthens the implementation of explainability and (non-judicial) contestability mechanisms.

\subsubsection{Recommendation 2: Combining indirect and direct control mechanisms.} 

In the interviews, a consensus emerged that contestability allows individuals to exercise control over AI usage by public institutions, and that it is based on information embodied by explainability. While individual action was seen as a way to assess the acceptability of specific decisions, collective action was considered more appropriate for challenging the system's overall suitability. Legal research suggests that the optimal governance scheme combines indirect control (by a regulation authority) and direct oversight (by decision subjects who appeal decisions and get redress, also called ``democratic control''~\cite{bogiatzis-gibbons_2024})~\cite{graham2003principles, smuha2021beyond, wachter2024limitations}. We find that this direct oversight by decision subjects through contestability mechanisms is both (i) considered positively by AI regulation experts and (ii) not very familiar to them when it comes to concrete ways to implement it. The equilibrium between indirect and direct control over public AI should thus be reconsidered, e.g., by giving more place to direct control using non-judicial contestation means. This can be supported by providing explanations that disclose the purpose of an AI system's development and deployment (normative) rather than merely describing its workings (descriptive)~\cite{bogiatzis-gibbons_2024}. Public institutions should thus ensure that provided explanations are relevant and aligned with the recipient's goals and level of knowledge~\cite{cobbe2021reviewable, norval2022disclosure} by considering which kind of contestation they enable~\cite{selbst_meaningful_2017}. Further, public institutions can engage with the regulation ecosystem (e.g., AI Office, standardization bodies) to receive support in the alignment between the regulatory intent of explainability and contestability and their implementation~\cite{prifti_regulation_2024}.  

\subsubsection{Recommendation 3: Ensuring the availability of non-judicial contestation channels.} The majority of the interviewees were familiar with judicial means of contestation and unfamiliar with non-judicial ones, which explains why this aspect of contestability remained overlooked in regulatory initiatives until now~\cite{kaminski2021right}. We argue that public institutions should adopt a more holistic approach to contestability that goes beyond ``complying to the letter'' to improve trust and acceptability. To this end, non-judicial means to implement contestability could be better leveraged in legal instruments guiding the implementation of AI regulations, especially as standards are developing into means of judicial control~\cite{schapel2013, gornet_european_2024}. While policy needs to decide \textit{what} can be contested, \textit{who} can contest, \textit{who} is accountable, and \textit{which types of review} should be put in place~\cite{lyons2022_conceptualising}, 
internal AI governance systems should ensure ways for non-judicial contestation. These include tools for scrutiny, annual assessments, or differential treatment (i.e., room to negotiate decisions between decision subjects and operators)~\cite{alfrink2024envisioning}. Such mechanisms should ensure that decision subjects are given an opportunity to understand their situation~\cite{lyons_whats_2022} and can articulate their voice in the process~\cite{yurrita_disentangling_2023}. Public institutions can 
benefit from engaging with design and HCI researchers regarding the implementation of contestability in AI system design.

\section{Limitations}
\label{sec:limitations}

Like any research, this study had limitations. The single 30-minute card sort, despite being refined through four pilot studies, likely influenced the depth of participants' exploration and classification of the cards. Further, the citation graph presented was a simplified subset of literature on explainability and contestability and did not capture the whole research landscape, but served as a prompt for participants to share their interdisciplinary experiences. Lastly, the focus on an EU context for recruitment and analysis of regulation and implementation procedures excluded comparisons with other jurisdictions, which is a fruitful direction for future research.  

\section{Conclusion}
\label{sec:conclusion}

In this work, we conceptualized explainability and contestability and their implementation by interviewing 14 interdisciplinary experts on AI regulation. We provided distinctions to facilitate implementation of these principles, including between normative and descriptive explanations, individual and collective contestation action, and judicial and non-judicial contestation channels. Key challenges include the preservation of the regulation's spirit, the responsibility for interpreting regulatory provisions, and the collaboration between disciplines. Based on these findings, we recommend i) strengthening the intersection between both principles in policy and governance, ii) considering non-judicial channels of contestation to improve trust, and iii) employing early-stage deliberations in the development of public AI systems to improve acceptability and avoid excessive contestation. With this work, we aim to inform research and policy efforts that leverage explainability and contestability in the development of trustworthy public AI systems. 

\section*{Acknowledgments}
This work has been funded by the Vienna Science and Technology Fund (WWTF) [10.47379/ICT20058] as well as [10.47379/ICT20065], and was made possible by the support of the French Embassy in Austria.

\section*{AI Usage Statement}
During the final preparation of this manuscript, we utilized three AI-assisted tools for copy-editing: Anthropic's Claude, Writefull, and Grammarly. These tools were used solely to improve clarity and readability without altering the paper's intellectual content, methodology, or findings.

\printbibliography

\section*{Appendix A: Card sort material and citation graph}
\addcontentsline{toc}{section}{Appendix A: Card sort material and citation graph}

\begin{figure}[htb]
    \centering
    \includegraphics[width=1.0\linewidth]{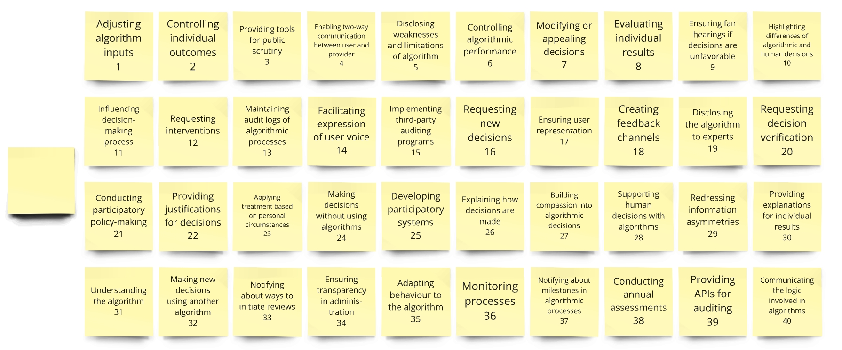}
    \caption[Card sort material]{\textbf{Card sort material.} The image depicts the full selection of 40 cards with explainability and contestability mechanisms. Participants received these cards and were asked to sort them into self-defined categories. Numbers were assigned at random, serving as IDs. New items could be added using the stack of empty cards.}
    \label{fig:card_sort}
\end{figure}

\begin{figure}
    \centering
    \includegraphics[width=1.0\linewidth]{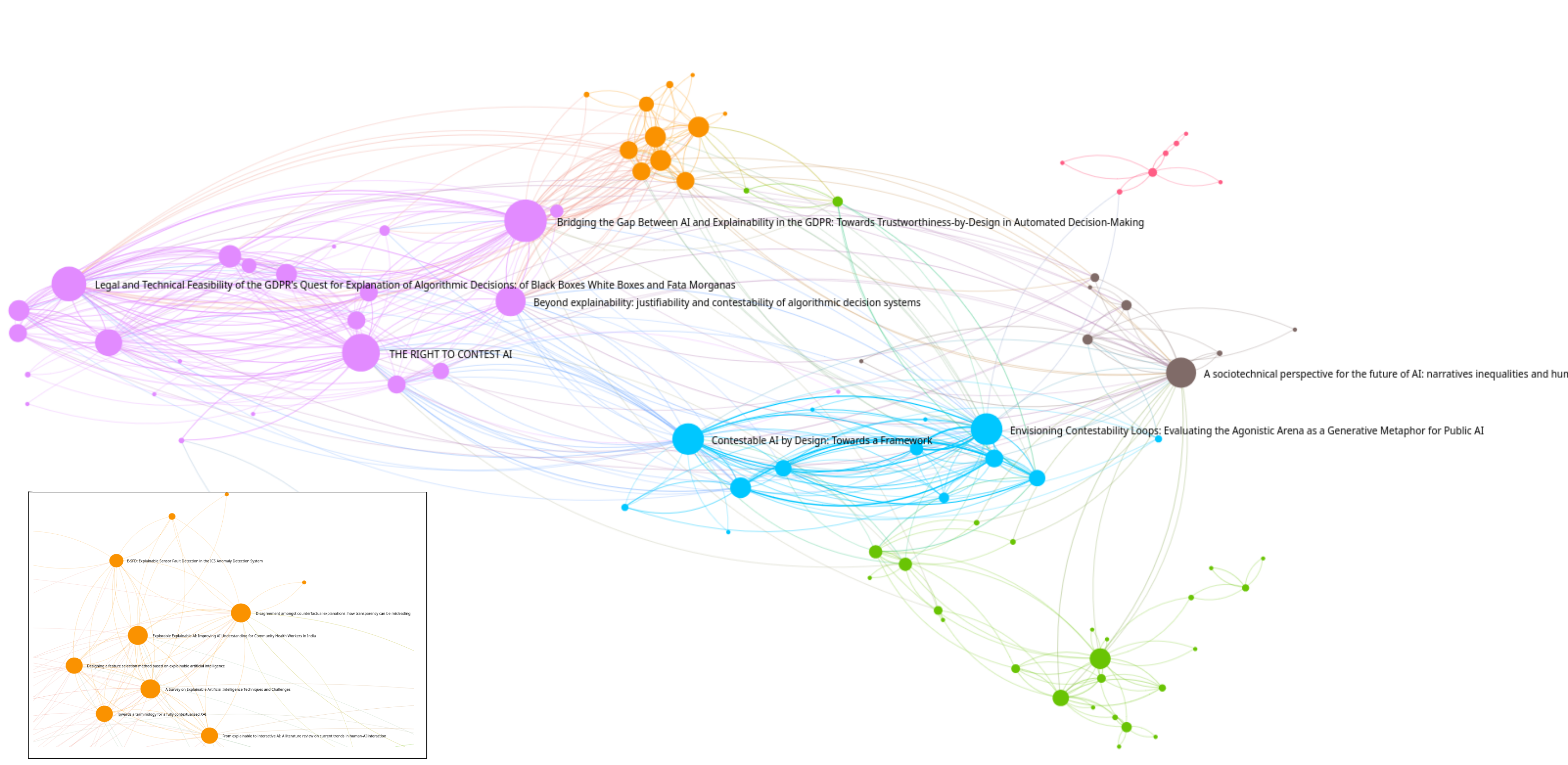}
    \caption{\textbf{Overview of the citation graph.} The image shows the citation graph used in the study to elicit participants' reflections on the research landscape. The graph is the largest connected component of the co-reference graph related to contestability and AI, and it also includes references on the `right to explanation'. The detail shows a zoom on the ``explainable AI'' cluster in the network.}
    \label{fig:literature-graph}
\end{figure}

\end{document}